\documentclass[notitlepage,floats,aps,nofootinbib,preprintnumbers,superscriptaddress,twocolumn,prd,10pt,balancelastpage,longbibliography]{revtex4-1}
\usepackage{amsmath,amssymb,amsfonts}
\usepackage{mathtools}
\usepackage{hyperref}
\hypersetup{colorlinks,citecolor=blue,urlcolor=blue,linkcolor=blue}
\usepackage{graphicx}
\usepackage{xcolor}
\usepackage{subfigure}
\usepackage{nicefrac}
\usepackage{mathrsfs}
\usepackage{mdframed}
\usepackage{units}
\usepackage{slashed}
\usepackage{wrapfig}

\newcommand{\gs}{g_\star}
\newcommand{\gss}{g_{\star s}}
\newcommand{\Trh}{T_\text{rh}}

\def\beq{\begin{equation}\begin{aligned}}
\def\eeq{\end{aligned}\end{equation}}

\begin{document}
%%%%%%%%%%%%%%%%%%%%%%%%%%%%%%%%%%%%%%%%%%%%%%
\title{Minimal Freeze-in Dark Matter:\\Reviving electroweak doublet dark matter with Boltzmann suppressed freeze-in}

\author{Nicol\'as Bernal}
\affiliation{New York University Abu Dhabi, PO Box 129188, Saadiyat Island, Abu Dhabi, UAE.}
\author{Sagnik Mukherjee}
\author{James Unwin}
\affiliation{Department of Physics,  University of Illinois Chicago, Chicago, IL 60607, USA.}

%%%%%%%%%%%%%%%%%%%%%%%%%%%%%%%%%%%%%%%%%%%%%%
\begin{abstract} 
%%%%%%%%%%%%%%%%%%%%%%%%%%%%%%%%%%%%%%%%%%%%%%
Dark matter communicating with the Standard Model solely via electroweak interactions provides a compelling picture. However, thermal freeze-out of electroweak doublet dark matter is generically strongly excluded by direct detection. We show that SU(2)${}_L$ doublet fermion dark matter evades direct detection if its mass exceeds $10^{10}$~GeV. If the neutral Dirac fermion is split into a pseudo-Dirac pair (via high dimension operator) this limit can be relaxed to 300~GeV.  Provided the dark matter mass is above the reheat temperature of the Universe, the production rate never exceeds the Hubble rate in cases of interest,  thus the dark matter never thermalizes. We apply constraints from direct detection (e.g.~LZ) and consider the discovery potential of Darwin. This scenario presents the most minimal model  of freeze-in dark matter, and is both elegant and highly predictive.
%%%%%%%%%%%%%%%%%%%%%%%%%%%%%%%%%%%%%%%%%%%%%%
\end{abstract}
%%%%%%%%%%%%%%%%%%%%%%%%%%%%%%%%%%%%%%%%%%%%%%
\maketitle

%%%%%%%%%%%%%%%%%%%%%%%%%%%%%%%%%%%%%%%%%%%%%%
\section{Introduction}
\vspace{-2mm}
%%%%%%%%%%%%%%%%%%%%%%%%%%%%%%%%%%%%%%%%%%%%%%
The most compelling dark matter models are those that only minimally extend the Standard Model. Three classic realizations are the Higgs portal,  dark matter with electroweak interactions, and (non-thermal)  gravitationally coupled dark matter. Notably, thermal freeze-out of dark matter via the Higgs or electroweak gauge bosons is generically excluded \cite{Goodman:1984dc, Escudero:2016gzx} (with some caveats, e.g. \cite{Tucker-Smith:2001myb, Davoudiasl:2015vba, Chanda:2019xyl}).  It was recently highlighted that Boltzmann suppressed freeze-in of scalar $\phi$ or fermion $\psi$ dark matter via the Higgs portals $|\phi|^2|H|^2$ \cite{Cosme:2023xpa} or $\bar\psi\psi|H|^2$ \cite{Bernal:2025fcl} leads to viable models, evading experimental constraints. Here we study Boltzmann-suppressed freeze-in of dark matter via electroweak gauge bosons, focusing on the realization in which dark matter is a fermion doublet of SU(2)${}_L$ with an electromagnetically neutral  component. 

For dark matter to have electroweak interactions it must be a non-trivial representation of SU(2)${}_L$. While different representations have been considered in the literature (see e.g.~\cite{Cirelli:2005uq}), there are no known examples of fundamental matter fields transforming in any representation other than the singlet or fundamental. For this reason, we focus on electroweak doublet dark matter. In this case, the Standard Model electroweak bosons can be the sole mediators between dark matter and the Standard Model, making the model highly predictive. Indeed, if the relic abundance of electroweak doublet dark matter is set by thermal freeze-out its mass is required to be TeV scale  (assuming standard cosmology \cite{Chanda:2019xyl}). This scenario is generically strongly excluded by direct detection \cite{Caldwell:1988su, LZ:2024zvo}. As we show, direct detection limits require the mass of doublet dark matter to be very heavy. 

We show that SU(2)${}_L$ doublet fermion dark matter is a viable possibility if its mass is heavier than $10^{10}$~GeV. This mass bound can be relaxed to 300~GeV if the neutral dark matter state is pseudo-Dirac. In the pseudo-Dirac case, the tree-level spin-independent (SI) direct detection cross-section can be forbidden; however, there is an irremovable loop-induced SI scattering cross-section. We also require that the dark matter mass exceeds the reheat temperature of the Universe: $m_{\rm DM}>\Trh$. In this case, the production rate typically never exceeds $H$, so the doublet never thermalizes.

Notably, while for  freeze-out dark matter the mass of  dark matter is bounded by partial-wave unitary constraints to be $\lesssim 100$~TeV \cite{Griest:1989wd}, there is no such upper bound for freeze-in dark matter, and (much) heavier dark matter is perfectly consistent. Moreover, as we demonstrate, for the  dark matter models studied here the mass range $10^{10}$ -- $10^{13}$ GeV has the potential to be discovered at forthcoming direct detection experiments.

%%%%%%%%%%%%%%%%%%%%%%%%%%%%%%%%%%%%%%%%%%%%%%%%%%%%%%%
\vspace{-2mm}
\section{Electroweak Doublet Dark~Matter}
\vspace{-2mm}
%%%%%%%%%%%%%%%%%%%%%%%%%%%%%%%%%%%%%%%%%%%%%%%%%%%%%%%
Non-trivial representations of SU(2)$_{L}$ are multi-component objects. Each component has a different electromagnetic (EM) charge after electroweak symmetry breaking (EWSB) given by $ Q = T_3 + Y $. For an $\mathrm{SU}(2)_L$ doublet, these components are labeled by the third component of their isospin $T_3 = \pm \frac{1}{2}$. To avoid coupling to photons, the EM charge must satisfy $ Q = 0 $, which implies $T_3 =- Y$. Thus, for a doublet to have an electromagnetically neutral component after EWSB, the hypercharge of the doublet must be $Y_\chi=\pm \frac{1}{2}$. 

We focus on fermion dark matter since scalar dark matter suffers from a fine-tuning `hierarchy' problem and also permits a model-independent renormalizable Higgs portal interaction, which makes it less predictive. For fermion dark matter, the minimal realization requires a pair of Weyl fermions $\chi_1$ and $\chi_2$ which are SU(2)${}_L$ doublets (to avoid the Witten anomaly \cite{Witten:1982fp}). While the Weyl fermions could have differing hypercharges (or other quantum numbers) unless one adds additional states, anomaly cancelation constrains  $Y[\chi_1]$ and $Y[\chi_2]$. 

The minimal anomaly-free construction involves two left-handed (LH) doublets with opposite hypercharge $\pm \frac{1}{2}$. This model has a large tree-level vector coupling to the $Z$ and thus leads to a large SI scattering cross-section. Interestingly, if one introduces a modest mass splitting between the two EM neutral degrees of freedom, then the tree level SI cross-section can be removed \cite{Tucker-Smith:2001myb}, which ameliorates these limits. We will examine this variant following the minimal model.

Explicitly, in the minimal model one supplements the Standard Model with two LH doublets of opposite hypercharge: $Y[\chi_1] = +\frac{1}{2}$ and $Y[\chi_2] = -\frac{1}{2}$. Each of the doublets can be written as follows with the EM charges shown as superscripts
\beq
    \chi_1=\begin{pmatrix}\chi_1^+\\ \chi_1^0\end{pmatrix},\quad
    \chi_2=\begin{pmatrix}\chi_2^0\\ \chi_2^-\end{pmatrix},
\eeq
and the Lagrangian is given by
\beq
    \mathcal L_{\chi} = i \chi_1^\dagger \bar\sigma^\mu D_\mu \chi_1 + i \chi_2^\dagger \bar\sigma^\mu D_\mu \chi_2 - \Big[ M \epsilon^{ab} \chi_{1 a}\chi_{2 b} + {\rm H.c.}\Big],
\eeq
where $D_\mu$ is the ordinary electroweak covariant derivative. Note that  $\epsilon^{ab}\chi_{1 a}\chi_{2 b}=\chi_1^+\chi_2^- - \chi_1^0\chi_2^0$, since we have $\epsilon^{12}=-\epsilon^{21}=1$. After EWSB, one can assemble the Weyl spinors into neutral  $\Psi^0$ and charged  $\Psi^\pm$  fields
\beq
    \Psi^0 \equiv \begin{pmatrix}\chi_2^0 \\ \overline{\chi}_1^0\end{pmatrix},\qquad
    \Psi^+ \equiv \begin{pmatrix}\chi_1^+ \\ \overline{\chi}_2^-\end{pmatrix}.
\eeq

The Lagrangian can be rewritten in terms of $\Psi$ with $\mathcal L_\Psi=\mathcal L_{\rm int}+\mathcal L_{M}$ with 
{\small 
\beq
    \mathcal L_{\rm int} =& \frac{g_2}{\cos\theta_W} Z_\mu \left[\left(\frac{1}{2} - \sin^2\theta_W\right)\bar\Psi^+ \gamma^\mu \Psi^+ +  \frac{1}{2} \bar\Psi^0 \gamma^\mu \Psi^0 \right] \\
    &+ \frac{g_2}{\sqrt2} \Big(\bar\Psi^+ \gamma^\mu \Psi^0  W_\mu^+  +  \bar\Psi^0 \gamma^\mu \Psi^+  W_\mu^- \Big) + e A_\mu \bar\Psi^+ \gamma^\mu \Psi^+.
    \label{eq:4}
\eeq}
Section \ref{ApA} explains the coupling structure. Further, this construction permits a Dirac mass
\beq
    \mathcal L_{M} = - M \bar\Psi^0 \Psi^0 - M \bar\Psi^+ \Psi^+ .
\label{LM}\eeq
Thus for fermion dark matter the second Weyl field is not only needed for anomaly cancelation but provides the minimal manner of giving an appropriately large mass to the dark matter. At tree level, the charged and neutral Dirac fermions are degenerate. After EWSB, radiative corrections (for $M\gg m_Z$) imply \cite{Cirelli:2005uq}
\beq\label{350}
    \Delta  \equiv m_{\Psi^+}-m_{\Psi^0} \approx 350~{\rm MeV}.
\eeq
If the states carry a conserved charge (e.g.~$\mathbb{Z}_2:$  $\chi_i\rightarrow -\chi_i$) then $\Psi^0$  will be stable and is a dark matter candidate. 

%%%%%%%%%%%%%%%%%%%%%%%%%%%%%%%%%%%%%%%%
\section{Electroweak doublet $\boldsymbol{Z}$ couplings}
\label{ApA}
%%%%%%%%%%%%%%%%%%%%%%%%%%%%%%%%%%%%%%%%
%We recall that a pair of fermions transforming in the doublet representation of SU(2)${}_L$ provides an anomaly-free extension of the Standard Model which yields a potential dark matter candidate for a pair of LH Weyl fermions with opposite hypercharge $Y=\pm 1/2$. The restriction to $Y=\pm 1/2$ ensures that there is an electromagnetically neutral state after EWSB. 

As outlined above, after EWSB one can combine the components of these fields into a neutral Dirac fermion $\Psi^0$ and a charged Dirac fermion $\Psi^\pm$.
In this section, we derive the couplings of the new fermions following from their representations and hypercharge. For a Dirac fermion $\Psi$ with electric charge $Q$ and weak isospin eigenvalues $T_3^{L}$ and $T_3^{R}$, for its left- and right-handed components, the vector and axial-vector couplings of these states to the $Z$ are given by
\beq
    \label{qqq}
    \hat g_V &= \frac{e}{2 \sin\theta_W \cos\theta_W} \Big(T_3^{L} + T_3^{R} - 2 Q \sin^2\theta_W\Big), \\ 
    \hat g_A &= \frac{e}{2 \sin\theta_W \cos\theta_W} \Big(T_3^{L} - T_3^{R}\Big),
\eeq
with $e = g_2  \sin\theta_W$. We  add a hat to the coupling to show that the coupling is specialized to the $Z$. This corresponds to an interaction structure of the form
\beq
    \mathcal L \supset Z_\mu \bar\Psi\gamma^\mu\big(\hat g_V-\hat g_A\gamma^5\big)\Psi .
\eeq
For the charged state $\Psi^+$ one has $T_3^{L}=T_3^{R}=1/2$ and $Q=+1$, giving
\beq \label{www1}
    \hat g_V^{\Psi^+} &=\frac{e}{2\sin\theta_W\cos\theta_W}\left(1 - 2 \sin^2\theta_W\right), \\
    \hat g_A^{\Psi^+} &=0 ,
\eeq
similarly, for the neutral state $\Psi^0$ its coupling to $Z$ is given by
\beq \label{www2}
    \hat  g_V^{\Psi^0} &= \frac{e}{2\sin\theta_W \cos\theta_W}, \hspace{12mm} \hat g_A^{\Psi^0} = 0 .
\eeq
Thus we observe that both the neutral and charged components have a non-zero tree-level vector coupling and a vanishing axial vector coupling.

Additionally, the quark couplings come from eq.~\eqref{qqq}, with $T_3^{R}=0$ for Standard Model fermions, thus
\beq\label{q}
    \hat g_V^u &= \frac{e}{2\sin\theta_W \cos\theta_W}\Big(\frac{1}{2} - \frac{4}{3}\sin^2\theta_W \Big), \\
    \hat g_A^{u} &= \frac{e}{4\sin\theta_W \cos\theta_W} , \\
    \hat g_V^{d}&=\frac{e}{2\sin\theta_W \cos\theta_W}\big(-\frac{1}{2} + \frac{2}{3}\sin^2\theta_W \big),\\
    \hat g_A^{d} &= -\frac{e}{4\sin\theta_W \cos\theta_W} .
\eeq

We will take $\sin^2\theta_W|_{\mu=m_Z}\approx 0.231$ here; this is the $Z$ pole value, however, the parameter does not significantly change via running to  intermediate scales. Evaluating $\sin^2\theta_W(\mu) = g_Y^2(\mu)/(g_Y^2(\mu)+g_2^2(\mu))$ assuming Standard Model running it is seen that $\sin^2\theta_W\approx 0.36\pm0.06$ for $\mu\sim 10^{12\pm4}$ GeV \cite{ParticleDataGroup:2024cfk}. Thus, the running of $\sin^2\theta_W$ will not significantly impact our calculations.  Similarly, all our calculations are at tree-level. It is expected that results will only vary at the $\mathcal{O}(10\%)$ at greater precision.

%%%%%%%%%%%%%%%%%%%%%%%%%%%%%%%%%%%%%%%%%%%%%%%%%%%%%%%
\vspace{-2mm}
\section{Boltzmann Suppressed Freeze-in}
\vspace{-2mm}
%%%%%%%%%%%%%%%%%%%%%%%%%%%%%%%%%%%%%%%%%%%%%%%%%%%%%%%
Dark matter freeze-in \cite{Hall:2009bx, Elahi:2014fsa} presents a distinctive alternative the to classic WIMP freeze-out picture. Boltzmann suppressed freeze-in, first outlined in Giudice-Kolb-Riotto (2000) \cite{Giudice:2000ex}, has recently received renewed attention, e.g. \cite{Cosme:2023xpa,  Cosme:2024ndc, Okada:2021uqk, Koivunen:2024vhr, Arcadi:2024wwg, Boddy:2024vgt, Arcadi:2024obp, Bernal:2024ndy, Lee:2024wes, Belanger:2024yoj, Khan:2025keb, Bernal:2025osg}. These scenarios consider the very conceivable possibility that the dark matter mass is higher than the maximum temperature of the thermal bath: $m_{\rm DM}\gg \Trh$. In this case the dark matter production cross-section is exponentially suppressed 
\beq
    \langle \sigma v\rangle \propto \exp(-2\, m_{\rm DM}/T).
\eeq

The cross-section for dark matter production via the process $q\bar q' \to \Psi^+\Psi^-, \bar\Psi^0 \Psi^0, \bar{\Psi}^0\Psi^+$ involving an electroweak boson  (derived in Appendices \ref{ApB} \& \ref{ApD}) is given at leading order by
\begin{equation}
    \sigma_{q\bar q'\to \Psi\Psi}(\beta) = \frac{\beta}{24\pi s} \Big((g_V^{q})^2 + (g_A^{q})^2\Big) (g_V^{\Psi})^2\left(1-\frac{\beta^2}{3}\right),
\end{equation}
where $\beta(s) \equiv \sqrt{1-4(m_\Psi^2/s)}$. Note that the axial vector contribution occurs at $\mathcal{O}(\beta^3)$. The general form of $\sigma_{q\bar q'\to \Psi\Psi}$ is useful and holds for arbitrary $q,q'$ initial states. This form can be specialized to different initial and final states by specifying the couplings. 

Turning to the Boltzmann equation
\beq \label{eq:BE}
    \dot n_\Psi + 3  H  n_\Psi = \gamma(T)\,.
\eeq
Neglecting the quark masses, the reaction density $\gamma$ for $q\bar q'\to\Psi^+\Psi^-$, $\bar{\Psi}^0\Psi^0$, $\bar{\Psi}^0\Psi^+$ (the latter arises from $u\bar{d}$ initial state, or similar, via $W$ exchange) is given by
\beq
  \gamma = \frac{T}{32\pi^4} \sum_{q,q'}\int_{4m_\Psi^2}^{\infty} {\rm d}s \Big[ (s-4m_\Psi^2)\sqrt{s} \sigma_{q\bar q'\to \Psi\Psi}  \Big] K_1 \left(\frac{\sqrt{s}}{T}\right) ,
  \label{eq:gamma}
\eeq
where $K_1$ is the modified Bessel function.

In the limit $m_\Psi \gg T$, the integral in eq.~\eqref{eq:gamma} is dominated just above threshold for which $s\simeq 4 m_\Psi^2$ and $\beta\ll1$. Thus, the near-threshold form of the cross-section is 
\beq\label{abc}
    \sigma_{q \bar q' \rightarrow \Psi\Psi}(s) \simeq \frac{1}{24 \pi  s}  (\mathcal C_\gamma^{qq'} + \mathcal C_Z^{qq'} + \mathcal C_W^{qq'})  \beta\,,
\eeq
where we have collected the combination of couplings into
\beq
    \mathcal C_i^{qq'}  \equiv \Big((g_V^{q})^2+(g_A^{q})^2\Big)(g_V^{\Psi})^2.
    \label{C}
\eeq
In Appendices \ref{ApC} \& \ref{ApD} we derive numerical values of $\mathcal C_Z$, $\mathcal C_\gamma$, and $\mathcal C_W$ for each case of interest. 

Further, as derived in Appendix \ref{ApE}, the reaction density (given eq.~\eqref{eq:gamma}) evolves according to
\beq
    \gamma(T) \simeq\sum_{q,q'}\left( \frac{ \mathcal C_\gamma^{qq'} + \mathcal C_Z^{qq'} + \mathcal C_W^{qq'}}{256 \pi^4} \right) T^4\,  e^{-2m_\Psi/T}.
    \label{abs}
\eeq
This can be specialized to the neutral or charged components by the appropriate choice of $\mathcal C_i$.

We then use $\gamma(T)$ to calculate the dark matter freeze-in yields $Y \equiv n_\Psi/s$ arising from the Boltzmann equation 
\beq \label{eq:yield}
    \frac{dY}{dT} =  - \frac{\gamma(T)}{s  H  T} ,
\eeq
where $s(T) = \frac{2\pi^2}{45}\gss T^3$ and $ H(T) = \sqrt{\frac{\pi^2 \gs}{90}}   \frac{T^2}{M_{\rm P}}$ in terms of the relativistic degrees of freedom $\gs$, the effective entropy degrees of freedom $\gss$ and the Planck mass $M_{\rm P}$. Substituting the reaction density from eq.~\eqref{abs} gives
\beq
    \frac{dY}{dT} = - \frac{135\sqrt{10}}{512 \pi^7} \sum_{q,q'}\left( \frac{\mathcal C_\gamma^{qq'} + \mathcal C_Z^{qq'} + \mathcal C_W^{qq'}}{\gss\sqrt{\gs}}  \right)\frac{M_{\rm P}}{T^2}  e^{-\frac{2m_\Psi}{T}}.
\eeq

%%%%%%%%%%%%%%%%%%%%%%%%%%%%%%%%%%%%%%%%%%%%%%%%%
\begin{figure}[t!]
    \def\sepf{1}
    \centering
    \includegraphics[width=\sepf\columnwidth]{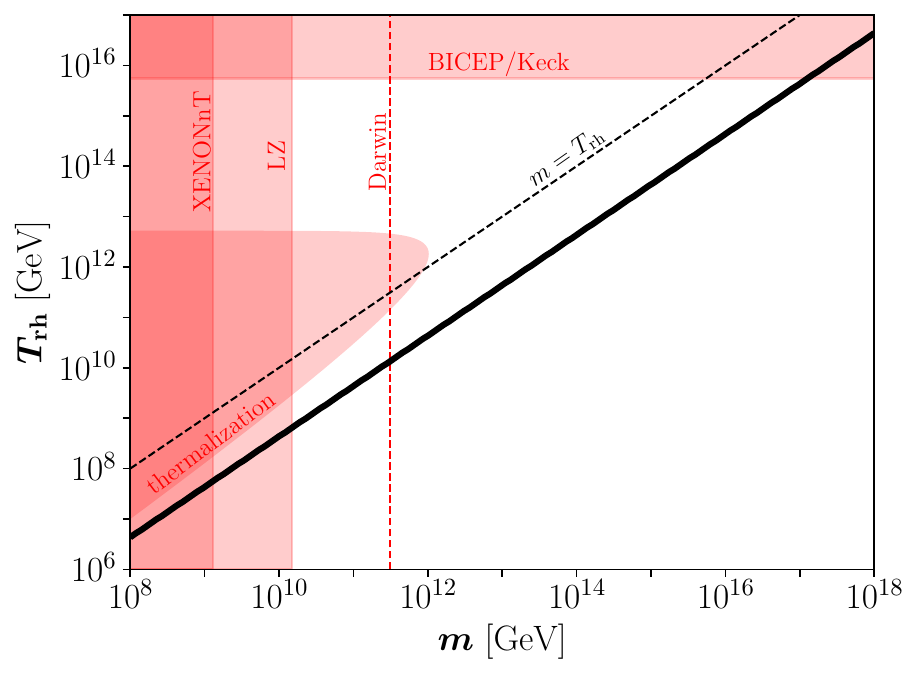}
    \vspace*{-3mm}
    \caption{The solid line is the  reheat temperature $\Trh$ which gives the observed dark matter relic density for $\Psi^0$ of mass $m$ assuming instantaneous reheating. The vertical red areas mark exclusion from direct detection experiments, and the projection by Darwin. The black dashed line indicates $\Trh = m$, above which freeze-in is no longer Boltzmann suppressed. The red shaded  ``thermalization'' region indicates parameter values for which $\Psi^0$ would enter equilibrium with the Standard Model.
    \label{fig:RD}}
    \vspace*{-2mm}
\end{figure} 
%%%%%%%%%%%%%%%%%%%%%%%%%%%%%%%%%%%%%%%%% 

Assuming instantaneous reheating\footnote{Boltzmann suppressed freeze-in of this model for non-instantaneous reheating is studied in the companion paper \cite{Bernal:2026vbg}, as well as a discussion of gravitational production (which is found to be always sub-leading).} to temperature $\Trh$, negligible initial $\Psi$ abundances, and taking $m_\Psi\gg \Trh$, the freeze-in yield is
\begin{align}
    Y_{\rm FI} &= \frac{135\sqrt{10}}{512 \pi^7} \sum_{q,q'}\frac{\mathcal C_\gamma^{qq'} + \mathcal C_Z^{qq'} + \mathcal C_W^{qq'}}{\gss\sqrt{\gs}} M_{\rm P} \int_0^{\Trh} dT  \frac{e^{\frac{-2m_\Psi}{T}}}{T^2} \nonumber\\
    &\simeq \frac{135\sqrt{10}}{1024 \pi^7} \sum_{q,q'} \frac{\mathcal C_\gamma^{qq'} + \mathcal C_Z^{qq'} + \mathcal C_W^{qq'}}{\gss\sqrt{\gs}} \frac{M_{\rm P}}{m_\Psi} e^{-\frac{2m_\Psi}{\Trh}}.
  \label{eq:Y0}
\end{align}
Observe that the freeze-in yield scales as $Y_{\rm FI} \propto \exp(-2m_\Psi/\Trh)$, as expected in the Boltzmann suppressed regime. From eq.~\eqref{eq:Y0} one can calculate the yield $Y^0_{\rm FI}$ of $\Psi^0$ and $Y^\pm_{\rm FI}$ of $\Psi^\pm$ by substituting the correct $ \mathcal C_i$.

Since  $m_{\Psi^\pm} > m_{\Psi^0}$ (cf.~eq.~\eqref{350}), the $\Psi^{\pm}$ will decay to $\Psi^0$ thus contributing to the dark matter relic density. The $\Psi^\pm  \to  \Psi^0 \pi^\pm$ decay rate is \cite{Thomas:1998wy} (cf.~also \cite{Ibe:2012sx})
\beq\notag
    \Gamma_{\Psi^\pm} &\simeq\Gamma_{\pi^\pm \to \mu^\pm \nu_\mu} \frac{16 \Delta^{3}}{m_\pi m_\mu^{2}} \sqrt{1 - \frac{m_\pi^{2}}{\Delta^{2}}} \left(1 - \frac{m_\mu^{2}}{m_\pi^{2}}\right)^{-2}.
\eeq 
The quantities above are all Standard Model and thus we evaluate numerically to find
$\Gamma_{\Psi^\pm} \approx 5 \times 10^{-14}~{\rm GeV}$.
Thus, the lifetime is $\tau\simeq 10^{-11}$~s and for cosmological purposes $\Psi^\pm$ decay rapidly. It follows that the dark matter relic abundance $Y^0_{\infty}$ is the sum $Y^0_{\infty}=Y^0_{\rm FI} + Y^\pm_{\rm FI}$.

In Fig.~\ref{fig:RD} we show (black solid line) the relic abundance of SU(2)${}_L$ doublet dark matter of mass $m$ as a function of the reheat temperature $\Trh$ (assuming instantaneous reheating). The black dashed line in Fig.~\ref{fig:RD} indicates $\Trh=m$, above which freeze-in is no longer Boltzmann suppressed. The red shaded  ``thermalization'' region indicates parameter values for which the dark matter would enter equilibrium with the Standard Model thermal bath; we have demanded the production rate to be smaller than the Hubble expansion rate $H$, that is, $\gamma(\Trh) < n_\text{eq}(\Trh) H(\Trh)$, where $n_\text{eq}$ is the dark matter equilibrium number density. The inflection in the boundary curve of the thermalization region in Fig.~\ref{fig:RD} corresponds to $m = \Trh$ which separates the relativistic and non-relativistic regimes. Observe that thermalization does not constrain the solid line for the relevant mass range. In addition, we impose an upper bound on the reheating temperature $\Trh \lesssim 5.5\times 10^{15}$~GeV, this comes from the BICEP/Keck limit on the tensor-to-scalar ratio.

%%%%%%%%%%%%%%%%%%%%%%%%%%%%%%%%%%%%%%%%% 
\vspace{-2mm}
\section{Direct Detection}
\vspace{-2mm}
%%%%%%%%%%%%%%%%%%%%%%%%%%%%%%%%%%%%%%%%% 
We next apply experimental limits from XENONnT \cite{XENON:2025vwd} and LZ \cite{LZ:2024zvo}. The bounds are linearly extrapolated from those presented in  \cite{LZ:2024zvo, XENON:2025vwd}. This extrapolation is  generally considered valid since heavy dark matter with sufficiently large cross-sections will interact with direct detection experiments, however, the higher mass implies that fewer states are needed to comprise the relic density. Thus, the dark matter flux is diminished,  leading to a linear fall off in the constraints. This can be seen in LZ's dedicated  heavy dark matter analysis \cite{LZ:2024psa}. The spin-independent cross-section for $\Psi^0$ scattering on an individual nucleon $N = p, n$ is \cite{Essig:2007az}
\beq\label{eq:SI}
    \sigma_{\rm SI}^{(N)} = \frac{\mu_N^2}{\pi}\left(\frac{ g_V^{\Psi^0}  g_V^{N}}{m_Z^2}\right)^2 \sim
    \begin{cases}
        4.7\times 10^{-40}~\text{cm}^2 & N=n, \\[4pt]
        2.6\times 10^{-42}~\text{cm}^2 & N=p,
    \end{cases}
\eeq
where $g_V^p=2 g_V^{u}+ g_V^{d}$ and $g_V^n= g_V^{u}+2 g_V^{d},$ for $g_V^{u},g_V^{d}, g_V^\Psi$ in Section \ref{ApA}. The 2025 LZ limit is \cite{LZ:2024zvo} 
\beq
    \left.\sigma_{\rm SI}^{(n,p)}\right|_{m_{\rm DM}\gg m_Z} < 3\times 10^{-47} {\rm cm}^2 \left(\frac{m_{\rm DM}}{{\rm TeV}}\right),
\eeq
implying a lower mass bound $m_{\rm DM} > 1.4\times 10^{10}$~GeV.

%%%%%%%%%%%%%%%%%%%%%%%%%%%%%%%%%%%%%%%%%%%%%%%%%
\begin{figure}[t!]
    \def\sepf{1}
    \centering    \includegraphics[width=\sepf\columnwidth]{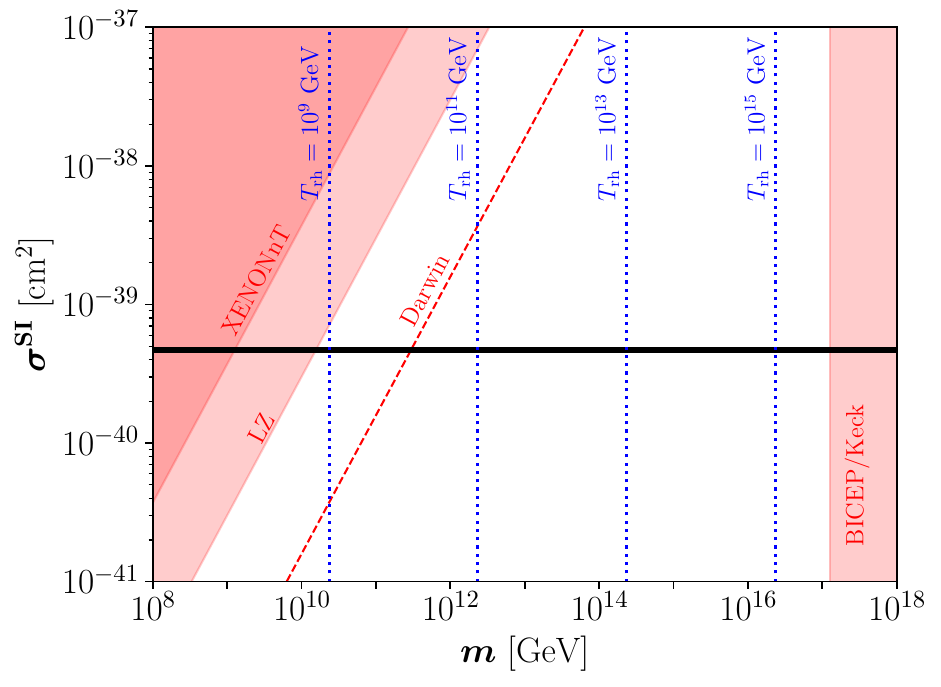}
    \vspace{-4mm}
    \caption{Constraints on Minimal Freeze-in Dark Matter. We apply the spin-independent limits from XENONnT \cite{XENON:2025vwd} and LZ \cite{LZ:2024zvo}. We also show the anticipated discovery reach of the proposed Darwin experiment \cite{DARWIN:2016hyl}. The parameter space below the `Darwin' line lies within the neutrino fog. For instantaneous reheating the dark matter mass uniquely determines the required $\Trh$ and we plot some characteristic contours. We show the cosmological limits (assuming single field inflationary reheating) from BICEP/Keck \cite{BICEP:2021xfz} which constrain $\Trh$.}
    \label{fig:SI}
    \vspace*{-3mm}
\end{figure} 
%%%%%%%%%%%%%%%%%%%%%%%%%%%%%%%%%%%%%%%%% 

In Fig.~\ref{fig:SI} we show the  SI limits on electroweak doublet dark matter. For instantaneous reheating, the dark matter mass uniquely determines the required $\Trh$, and these are shown as contours. We also show the cosmological limits from BICEP/Keck \cite{BICEP:2021xfz} that constrain $\Trh$ and this restricts $m \lesssim 10^{17}$ GeV. We find that indirect detection does not constrain the parameter space of interest; see Appendix \ref{ApF}. These mass bounds are also marked on Fig.~\ref{fig:RD} as shaded areas.

It is interesting to consider the discovery reach of the future Darwin experiment \cite{DARWIN:2016hyl}. Darwin probes deep into the available parameter space, although there remains a window between Darwin's SI reach and the BICEP/Keck limit. However, the parameter space below Darwin's reach lies in the neutrino fog\footnote{For $m \gg$ TeV the recoil spectrum is mass-independent (see e.g.~\cite{OHare:2020lva}). Since the dark matter flux scales as $1/m$, maintaining a fixed event rate relative to the (mass-independent) neutrino background requires $\sigma_{\rm SI}\propto m$, leading to a linear rise in the neutrino fog (i.e.~running parallel to the dark matter constraint).} \cite{DARWIN:2016hyl, OHare:2020lva} and thus will require novel experimental advancements to completely exclude these models. More optimistically, the Darwin experiment can discover electroweak doublet dark matter with mass in the range $10^{10}$ -- $10^{13}$ GeV.

%%%%%%%%%%%%%%%%%%%%%%%%%%%%%%%%%%%%%%%%%%%%%55
\vspace{-3mm}
\section{Pseudo Dirac Case}
\vspace{-3mm}
%%%%%%%%%%%%%%%%%%%%%%%%%%%%%%%%%%%%%%%%%%%%%55
Interestingly, higher-dimensional operators, generated by high-scale physics, may alter the expectations for direct detection. In the construction above, the neutral Dirac fermion $\Psi^0 \equiv(\chi_2^0, \overline{\chi}_1^0)$ has a tree-level vector coupling to the $Z$, which leads to a large SI scattering rate. The  tree-level $Z$ coupling can be removed by splitting the neutral Dirac fermion into two Majorana states (a pseudo-Dirac pair). This can be achieved via a small Majorana mass for the neutral components \cite{Tucker-Smith:2001myb}. A clean realization of this is via dimension-5 operators involving the Standard Model Higgs schematically of the form $\frac{1}{\Lambda}(\chi_1 H)^2$, where $\Lambda$ is the  mass scale of new physics. This is reminiscent of the Weinberg operator $(LH)^2$.

At low energy, one may parameterize the splitting via 
\vspace{-2pt}
\beq
    \mathcal L \supset - M \chi_1^0 \chi_2^0 -\frac{\delta_1}{2} \chi_1^0\chi_1^0 -\frac{\delta_2}{2} \chi_2^0\chi_2^0 +{\rm H.c.},
    \label{aa1}
\eeq
\vspace{-2pt}
where $M$ is the Dirac mass from eq.~(\ref{LM}) and $\delta_{1,2}\ll M$ are the induced Majorana masses. Then, the physical states are 
\beq
    \Psi_1 &\simeq \frac{1}{\sqrt 2}\left(\Psi^0 + (\Psi^0)^c\right),\\ 
    \Psi_2 &\simeq \frac{-i}{\sqrt 2}\left(\Psi^0 - (\Psi^0)^c\right),
\eeq
with split masses $m_{1,2} \simeq M \mp \frac{1}{2}\delta$ for $\delta\equiv \delta_1+\delta_2$. Being Majorana fermions the vector current vanishes, but an off-diagonal $Z$ coupling remains $\bar\Psi_2\gamma^\mu \Psi_1$. While inelastic scatterings $\Psi_1 N\to \Psi_2 N$ can be possible, for $\delta\gg  O(100)~{\rm keV}$ this is forbidden.

In the absence of inelastic scattering, the leading direct detection cross-section arises from EW loop-induced operators \cite{Hisano:2010ct}. The characteristic size for such loop-induced scattering cross-sections is $10^{-47}$~cm$^2$, roughly a factor $\mathcal{O}(10^{-7})$ below the tree-level case. The reduced SI cross-section significantly weakens the direct detection limits.\footnote{Since the mass splitting is generated by EWSB, freeze-in is unaffected  because with $m,T_{\rm rh}\gg m_Z$ the mass splitting vanishes.} Compared to direct detection limits, for $\sigma_{\rm SI}\sim10^{-47}~{\rm cm}^2$ the limit on the dark matter mass is $m\gtrsim 330$ GeV with Darwin probing up to 6 TeV. This requires a low $\Trh$ and allows an interesting alternative at the EW scale. Current collider and indirect detection bounds do not further constrain this model \cite{Panci:2024oqc}; indeed it has been suggested the 14 year Fermi-data may mildly favor TeV-scale doublet dark matter \cite{Dessert:2022evk}. Such a model is discoverable at the forthcoming CTAO-North experiment \cite{Abe:2025lci}. 

%%%%%%%%%%%%%%%%%%%%%%%%%%%%%%%%%%%%%%%%%%%%
\section{Concluding Remarks}
%%%%%%%%%%%%%%%%%%%%%%%%%%%%%%%%%%%%%%%%%%%%
The idea of dark matter communicating with the Standard Model only via electroweak interactions presents a compelling picture. We have shown that within the framework of Boltzmann suppressed freeze-in, electroweak doublet dark matter is consistent with current searches. We highlight that this model is arguably\footnote{Gravitationally coupled scalars $\Phi$ may seem minimal, but sub-$M_{\rm P}$ mass scalars must add new states to avoid fine-tuning, they also worsen the hierarchy problem of the Higgs due to $|H|^2|\Phi|^2$.} {\em the} most minimal dark matter freeze-in scenario imaginable in terms of field content introducing only a single new Weyl fermion pair and nothing else. Due to this minimality, the scenario presented is extremely predictive. The predictive strength of this setting  is partly due to our (not unreasonable) assumption of efficient inflationary reheating. For inefficient reheating or non-standard cosmology (different from matter-like prior to reheating; e.g.~kination domination, cf.~\cite{Allahverdi:2020bys}), then the maximum temperature of the Universe $T_{\rm max}$ and the equation of state of the early Universe $\omega$ will typically impact the freeze-in calculations (cf.~\cite{Bernal:2019mhf, Allahverdi:2020bys}). Moreover, it is interesting to consider freeze-in of SU(2)${}_L$ representations other than the doublet (for instance, the triplet or quintuplet representations). We study the impact of both the particle content and cosmological variations, and how they alter constraints and detection prospects, in a companion paper \cite{Bernal:2026vbg}. 

%\vspace{2mm}
\section*{\bf Acknowledgements}
 NB received grants PID2023-151418NB-I00 funded by MCIU/AEI/10.13039/ 501100011033/ FEDER 
and PID2022-139841NB-I00 
 of MICIU/AEI/10.13039/501100011033 and FEDER, UE.
JU is supported by NSF grant PHY-2209998.

%%%%%%%%%%%%%%%%%%%%%%%%%%%%%%%%%%%%%%%%
\appendix
%%%%%%%%%%%%%%%%%%%%%%%%%%%%%%%%%%%%%%%%
\bibliographystyle{apsrev4-1} 
\bibliography{biblio}
%%%%%%%%%%%%%%%%%%%%%%%%%%%%%%%%%%%%%%%%

\newpage
\onecolumngrid
%%%%%%%%%%%%%%%%%%%%%%%%%%%%%%%%%%%%%%%%
\appendix
\section*{Appendices}

In the main text, we showed that the correct dark matter relic density could be reproduced for electroweak doublet fermion dark matter while avoiding direct detection constraints for heavy dark matter within the Boltzmann suppressed freeze-in framework. In the following, we provide details that support the main text, specifically:
\begin{enumerate}
%    \item We identify the electroweak doublet $Z$ couplings;
    \item A calculation of $q\bar q \to \Psi^+ \Psi^-,~ \bar\Psi^0 \Psi^0$ cross-sections;
    \item A computation of the numerical value of the coupling combination $\mathcal C_Z$ and $\mathcal C_\gamma$;
    \item A calculation of $q\bar q' \to \Psi^\pm \Psi^0,~ \bar\Psi^0 \Psi^0$ cross-sections via $W^\pm$;
    \item A derivation of the reaction density $\gamma(T)$;
    \item A discussion of indirect detection prospects.
\end{enumerate}

%%%%%%%%%%%%%%%%%%%%%%%%%%%%%%%%%%%%%%%%
\subsection{Freeze-in production via photon and $\boldsymbol{Z}$-boson}
\label{ApB}
%%%%%%%%%%%%%%%%%%%%%%%%%%%%%%%%%%%%%%%%
In this appendix, we derive the production cross-section $q\bar q \to \Psi^+ \Psi^-,~ \bar\Psi^0 \Psi^0$. As we calculate the freeze-in production in the regime $T\gg m_Z$, it can be considered more intuitive to work in the $(\chi_1,\chi_2)$, corresponding to the symmetric phase on the basis $(B,W_3)$. However, the results of the calculations are basis independent and since at late time, after EWSB, the $\Psi^0$ states are identified as dark matter, we find it simpler to always work in the $(\Psi^0,\Psi^\pm)$ basis. In what follows, we neglect the quark masses and work in the limit $m_Z^2\ll m_\Psi^2$, $s$.

The process $  q(p_1)  \bar q(p_2) \to \Psi^+(k_1)  \Psi^-(k_2)$ leading to the production of a pair of fermions with mass $m_\Psi$ and momenta $k_1,k_2$ from two quarks with momenta $p_1,p_2$ via a vector boson arises from the interaction Lagrangian
\beq\label{L}
  \mathcal{L}_{\mathrm{int}} = V_\mu\left[\bar q \gamma^\mu\left(g_V^q - g_A^q \gamma^5\right) q + \bar \Psi \gamma^\mu\left(g_V^\Psi - g_A^\Psi \gamma^5\right)\Psi\right],
\eeq
with generic vector and axial-vector couplings $g_V^{q,\Psi}$, $g_A^{q,\Psi}$ given in eq.~\eqref{qqq}. Neglecting the mediator mass, so the propagator is $- i/s$  the tree-level amplitude is
\beq
  \mathcal{M} = \frac{-i}{s} \big[\bar v(p_2) \gamma^\mu \left(g_V^q - g_A^q \gamma^5\right) u(p_1)\big] \big[\bar u(k_1) \gamma_\mu \left(g_V^\Psi - g_A^\Psi \gamma^5\right) v(k_2)\big].
\eeq
Squaring the matrix element and summing over final-state spins and colors and averaging over initial-state spins and colors, it follows that $\overline{|\mathcal M|^2} \equiv \frac{1}{(2\times2)N_c^2} \sum_{\text{spins, colors}}|\mathcal M|^2 = \frac{1}{4N_c}\sum_{\text{spins}}|\mathcal M|^2 =\frac{1}{12} \mathrm{Tr}[\cdots](\frac{1}{s})^2 \mathrm{Tr}[\cdots]$,  which leads to
\beq
    \overline{|\mathcal{M}|^2} &= \frac{1}{12 s^2} {\rm Tr} \Big[ \slashed{p}_2 \gamma^\mu \left(g_V^q - g_A^q \gamma^5\right) \slashed{p}_1 \gamma^\nu \left(g_V^q - g_A^q \gamma^5\right) \Big] {\rm Tr} \Big[ (\slashed{k}_1 + m_\Psi)\gamma_\mu \left(g_V^\Psi - g_A^\Psi \gamma^5\right) (\slashed{k}_2 - m_\Psi)\gamma_\nu \left(g_V^\Psi - g_A^\Psi \gamma^5\right) \Big] \\[4pt]
    &= \frac{1}{12 s^2} \Big[ 4\big((g_V^{q})^2+(g_A^{q})^2\big) \left(p_1^\mu p_2^\nu + p_1^\nu p_2^\mu - g^{\mu\nu} p_1 \cdot p_2\right) \Big] \\
    &\qquad\times \Big[ 4\big((g_V^{\Psi})^2+(g_A^{\Psi})^2\big) \left(k_1^\mu k_2^\nu + k_1^\nu k_2^\mu - g^{\mu\nu}(k_1 \cdot k_2 - m_\Psi^2)\right) -4\big((g_V^{\Psi})^2-(g_A^{\Psi})^2\big)m_\Psi^2 g^{\mu\nu} \Big] \\[4pt]
    &= \frac{2}{3 s^2}\Bigg[ \Big((g_V^{q})^2+(g_A^{q})^2\Big) \Big( (g_V^{\Psi})^2\big(t^2+u^2+2m_\Psi^2 s\big) + (g_A^{\Psi})^2\big(t^2+u^2-2m_\Psi^2 s\big) \Big) +4  g_V^q g_A^q g_V^\Psi g_A^\Psi  s (t-u)\Bigg].
    \label{eq:2}
\eeq

In terms of the Mandelstam invariants,
\beq
    s=(p_1+p_2)^2,\qquad\qquad t=(p_1-k_1)^2,\qquad\qquad u=(p_1-k_2)^2,
\eeq
with $s+t+u=2m_\Psi^2$ for massless initial-state quarks. Defining
\beq
    \beta \equiv \sqrt{1-\frac{4m_\Psi^2}{s}}\, ,
\eeq
and $\vartheta$ as the angle between $\vec p_1$ and $\vec k_1$, one has
\beq
    t = m_\Psi^2-\frac{s}{2}\left(1-\beta\cos\vartheta\right),\qquad\qquad u = m_\Psi^2-\frac{s}{2}\left(1+\beta\cos\vartheta\right).
\eeq

The differential cross-section is
\beq
    \frac{d\sigma}{d\Omega} = \frac{1}{64 \pi^2\, s}\,  \frac{|\vec k|}{|\vec p|}\, \overline{|\mathcal{M}|^2} = \frac{\beta}{64 \pi^2\, s}\, \overline{|\mathcal{M}|^2}\,.
\eeq
Substituting eq.~\eqref{eq:2} and integrating the azimuthal angle to get a factor $2\pi$  gives
\beq\notag
    \frac{d\sigma}{d\cos\vartheta} &= \frac{\beta}{32\pi  s}  \frac{1}{3} \Bigg\{ \Big((g_V^{q})^2 + (g_A^{q})^2\Big)\Big[(g_V^{\Psi})^2 \left(1+\cos^2\vartheta + (1-\beta^2)\sin^2\vartheta\right) + (g_A^\Psi)^2 \beta^2(1+\cos^2\vartheta)\Big]
   + 4  g_V^q   g_A^q  g_V^\Psi  g_A^\Psi \beta\cos\vartheta \Bigg\},
\eeq
where we used $m_\Psi^2 = \frac{s}{4}(1-\beta^2)$ and we have integrated the azimuthal angle over $2\pi$. Integrating the differential cross-section over $\cos\vartheta \in [-1,1]$ leads to the cross-section
\beq
    \sigma_{q\bar q\to \Psi\Psi}(\beta) = \frac{\beta}{24\pi s} \Big((g_V^{q})^2 + (g_A^{q})^2\Big) \left[ (g_V^{\Psi})^2 \left(1-\frac{\beta^2}{3}\right) + (g_A^{\Psi})^2 \frac{2\beta^2}{3} \right].
    \label{yu}
\eeq
Note that when integrating over $\cos\vartheta$ the term proportional to $g_V^q  g_A^q  g_V^\Psi  g_A^\Psi$ vanishes. Thus, for the vector-like model (with $g_V\neq0$ and $g_A=0$) to leading order in $\beta$
\beq
    \sigma_{q\bar q\to \Psi\Psi}^{(Z)}(\beta) \simeq \frac{\beta}{24\pi s}  \Big((g_V^{q})^2 + (g_A^{q})^2\Big)  (g_V^{\Psi})^2.
    \label{eq:xsec2}
\eeq
Additionally, the quark couplings come from eq.~\eqref{q}, numerically these are approximately
\beq
    \hat g_V^u &\approx 0.07, \quad\quad
    \hat g_A^{u} &\approx 0.19, \quad\quad
    \hat g_V^{d}&\approx -0.13,\quad\quad
    \hat g_A^{d} &\approx -0.19,
\eeq
where we use $\sin^2\theta_W\approx 0.231$ to obtain the numerical values.

It remains to state explicitly the $Z$ mediated production cross-section for the vector-like case, which is given by
\beq
    \sigma_{q\bar q\to \bar{\Psi}^0\Psi^0}^{(Z)} &\simeq \frac{\pi\alpha^2}{24\sin^4\theta_W \cos^4\theta_W} \frac{\beta}{ s } \Big[\big(T_3^q-2Q_q \sin^2\theta_W\big)^2+(T_3^q)^2\Big],\\
    \sigma^{(Z)}_{q\bar q\to \Psi^+\Psi^-} &\simeq \frac{\pi\alpha^2}{24\sin^4\theta_W \cos^4\theta_W} \frac{\beta}{ s} \Big[\big(T_3^q-2Q_q \sin^2\theta_W\big)^2+(T_3^q)^2\Big] \Big[\big(1-2\sin^2\theta_W\big)^2\Big].
\eeq
For $T \ll m_\Psi$, production is dominated near-threshold, thus $s\simeq 4 m_\Psi^2$ and $\beta\ll1$.

The process $q  \bar q \to \bar\Psi^0   \Psi^0$ is only mediated by  the $Z$ boson, and in this case we use $g_{A,V}^{\Psi} = g_{A,V}^{0}$ in eq.~\eqref{eq:xsec2}. For the process $q  \bar q \to \Psi^+  \Psi^-$ there is both $Z$-mediated production (for which we use $g_{A,V}^{\Psi} = g_{A,V}^{\pm}$) and photon-mediated production contribution. We can specialize the general formula of eq.~\eqref{eq:xsec2} to the photon case by taking $g_A^q  = g_A^\Psi=0$ with $g_V^q = e  Q_q$, and $g_V^{\Psi\pm} = e$ (since $Q_\Psi=1$), leading to 
\beq
    \sigma^{(\gamma)}_{q\bar q\to \Psi^+\Psi^-}(s) = \frac{2 \pi  \alpha^2  Q_q^2}{3  s}  \beta \left(1 - \frac{\beta^2}{3}\right).
\eeq
Note that in the relativistic limit $\beta\rightarrow1$ this gives the standard Drell–Yan form.

It is interesting to consider the ratio of photon to $Z$ channels for the vector-like model; when referring to the $Z$ couplings, we will add a hat ($\hat g$) to indicate this and avoid confusion with the $\gamma$ couplings (which we immediately write in terms of $Q$). Recall from eqs.~\eqref{www1} \& \eqref{www2} that  $\hat g_V^{\Psi^\pm} = \frac{e}{2\sin\theta_W  \cos\theta_W} \left( 1 - 2 \sin^2\theta_W\right)$ and $\hat g_A^{\Psi^\pm}=0$  the ratio is 
\beq
    \frac{\sigma^{(Z)}_{q\bar q\to \Psi^+\Psi^-}}{\sigma^{(\gamma)}_{q\bar q\to \Psi^+\Psi^-}} \Bigg|_{T\ll m_\Psi} \simeq  
  \frac{(\hat g_V^{q})^2 + (\hat g_A^{q})^2 }{4e^2Q_q^2\sin^2\theta_W  \cos^2\theta_W} \left( 1 - 2 \sin^2\theta_W\right)^2 
\approx
    \begin{dcases}
        0.37 & ~~q=u,\\
        1.9 & ~~q=d.
    \end{dcases}
\eeq

%%%%%%%%%%%%%%%%%%%%%%%%%%%%%%%%%%%%%%%%%%%%%%%%%%%%%%%%%%%%
\subsection{Numerical value of the coupling combination} \label{ApC}
\vspace{-2mm}
%%%%%%%%%%%%%%%%%%%%%%%%%%%%%%%%%%%%%%%%%%%%%%%%%%%%%%%%%%%%
In this appendix, we calculate the value of the combination of couplings $\mathcal C$ defined in eq.~\eqref{C}. Recall that we define (compare to eq.~\eqref{eq:xsec2}) the $\mathcal C$ implicitly via
\beq
    \sigma(s) \simeq \frac{\beta}{24\pi s} (\mathcal C_\gamma^{qq'} + \mathcal C_Z^{qq'} + \mathcal C_W^{qq'}), \eeq
with
\beq
  \mathcal C_i  \equiv \Big((g_V^{q})^2+(g_A^{q})^2\Big)(g_V^{\Psi})^2.
\eeq

Evaluating the relevant combination of couplings for the neutral $ C_{Z}^0$ and charged $ C_{Z}^\pm$ components, one has
\beq
    \mathcal C_{Z}^0 &= \frac{e^2}{4\sin^2\theta_W \cos^2\theta_W}\Big((\hat g_V^{q})^2+(\hat g_A^{q})^2\Big),\\
    \mathcal C_{Z}^\pm &=\frac{e^2}{4\sin^2\theta_W \cos^2\theta_W}\Big((\hat g_V^{q})^2+(\hat g_A^{q})^2\Big) \big(1 - 2 \sin^2\theta_W\big)^2. 
\eeq

Then evaluating $\mathcal C$ we find for the $Z$ channel
\beq\label{C1}
    \mathcal C_{Z}^{0,q} & \approx
    \begin{dcases}
        5.5\times10^{-3} & q=u\\
       7.1\times10^{-3} & q=d
    \end{dcases}~; \hspace{1cm}
    \mathcal C_{Z}^{\pm,q} &\approx
    \begin{dcases}
        1.6\times 10^{-3}& ~~q=u\\
        2.1\times 10^{-3} & ~~q=d
  \end{dcases}~,
\eeq
 and for the charged component photon channel one has
\beq\label{C2}
    \mathcal C_{\gamma}^{\pm,q} =
    \begin{dcases}
        (2e^2/3)^2 \approx 4.3\times10^{-3} & q=u\\
        (e^2/3)^2 \approx 1.1\times10^{-3} & q=d
    \end{dcases}~.
\eeq

%%%%%%%%%%%%%%%%%%%%%%%%%%%%%%%%%%%%%%%%%%%%%%%%%%%%%%%%%%%%%%%%%%%%%%%%%%%
\subsection{Freeze-in production via $\boldsymbol{W}$-boson} \label{ApD}
%%%%%%%%%%%%%%%%%%%%%%%%%%%%%%%%%%%%%%%%%%%%%%%%%%%%%%%%%%%%%%%%%%%%%%%%%%%
In addition to the $Z$ and $\gamma$ mediated channels $q\bar q\to \Psi^+\Psi^-$ and $q\bar q\to \bar\Psi^0\Psi^0$, freeze-in for the electroweak doublet model can also proceed via `co-production' channels of the form
\beq
    u(p_1) \bar d(p_2)\to W^+ & \to \Psi^+(k_1) \Psi^0(k_2),\\
    d(p_1) \bar u(p_2)\to W^- & \to \Psi^-(k_1) \bar\Psi^0(k_2),
\eeq
and analogously for heavier quarks. The dominant charged-current co-production channels are those proportional to the largest CKM elements, specifically $u\bar d$, $c\bar s$, $t\bar b$ and conjugates. Other channels are CKM suppressed and lead to at most $\mathcal{O}(10\%)$ corrections. The relevant interactions are
\beq
    \mathcal L \supset \frac{g_2}{\sqrt2} W^+_\mu\left(\bar u \gamma^\mu P_L  V_{ud}  d\right) +\frac{g_2}{\sqrt2} W^+_\mu\left(\bar\Psi^+ \gamma^\mu P_L \Psi^0\right) +\text{H.c.},
\eeq
where $P_L=(1-\gamma^5)/2$ and $V_{ud}\approx 0.97$ is the CKM matrix element. The tree-level $u \bar d\to \Psi^+ \Psi^0$ matrix element is
\beq
    \mathcal M_W &= \frac{-i}{s} \left(\frac{g_2}{\sqrt2}V_{ud}\right)\left(\frac{g_2}{\sqrt2}\right) \Big[\bar v(p_2)\gamma^\mu P_L u(p_1)\Big] \Big[\bar u(k_1)\gamma_\mu P_L v(k_2)\Big]\\
    &= \frac{-i}{s}\, \big[\bar v(p_2)\gamma^\mu(g_V^{ud,W}-g_A^{ud,W}\gamma^5)u(p_1)\big]\, \big[\bar u(k_1)\gamma_\mu(g_V^{\Psi,W}-g_A^{\Psi,W}\gamma^5)v(k_2)\big],
    \label{eq:MW}
\eeq
where in the latter equation, we have matched the conventions of eq.~\eqref{L}, by identifying the effective vector and axial couplings:
\beq
    g_V^{ud,W}=g_A^{ud,W}=\frac{g_2}{2\sqrt2} V_{ud}, \qquad g_V^{\Psi,W}=g_A^{\Psi,W}=\frac{g_2}{2\sqrt2}.
    \label{eq:gW}
\eeq

Since $m_{\Psi^+}=m_{\Psi^0}$ (cf.~eq.~(\ref{350})) the kinematics are the same as the diagonal case (up to couplings), thus we reuse the general result of eq.~(\ref{yu})
\beq
    \sigma_{q\bar q'\to \Psi\Psi}(\beta) = \frac{\beta}{24\pi s} \Big((g_V^{q})^2 + (g_A^{q})^2\Big) \left[ (g_V^{\Psi})^2 \left(1-\frac{\beta^2}{3}\right) + (g_A^{\Psi})^2 \frac{2\beta^2}{3}\right].
\eeq
Substituting eq.~\eqref{eq:gW} gives the co-production cross sections
\beq
    \sigma^{(W)}_{u\bar d\to \Psi^+\Psi^0}(s)=\sigma^{(W)}_{d\bar u\to \Psi^-\bar\Psi^0}(s) = \frac{\beta}{24\pi s} \frac{g_2^4}{32} |V_{ud}|^2 \left(1+\frac{\beta^2}{3}\right).
\eeq
Taking the limit $\beta\ll1$, we write the cross-section in the form
\beq
    \sigma^{(W)}_{d\bar u\to \Psi^-\bar\Psi^0}(s) \Big|_{\beta\ll1} \simeq \frac{\beta}{24\pi s} \mathcal C_W^{qq'}.
\eeq
Noting that $C_W^{u\bar{d}}=C_W^{d\bar{u}}$ we drop the bars in the superscripts and for the first generation
\beq
    \mathcal C_W^{ud} = \frac{g_2^4}{32} |V_{ud}|^2 = \frac{1}{32} \frac{e^4}{\sin^4\theta_W} |V_{ud}|^2 = \frac{\pi^2\alpha^2}{2 \sin^4\theta_W} |V_{ud}|^2.
\eeq
For general up-type and down-type flavors $(u_i,d_j)$, the replacement is $|V_{ud}|^2  \to  |V_{ij}|^2$ to get the general $\mathcal{C}_W^{u_i d_j}\propto|V_{ij}|^2$. Using the values for the CKM matrix elements \cite{ParticleDataGroup:2024cfk}
\beq 
    |V_{ud}| \approx 0.97, \quad\quad\quad  |V_{cs}| \approx 0.98,  \quad\quad\quad |V_{tb}| \approx 1.0,
\eeq
numerically, the leading channels corresponding to $\mathcal C_W$ values of 
\beq
    \mathcal C_W^{ud} \approx 5.4\times 10^{-3},\quad\quad \mathcal C_W^{cs} \approx 5.4\times 10^{-3},\quad\quad \mathcal C_W^{tb} \approx 5.6\times 10^{-3}.
\eeq
Comparing with eqs.~\eqref{C1} \& \eqref{C2}, we see that these are comparable to the $Z$ and $\gamma$ mediator production. Accordingly, this leads to an $\mathcal O(1)$ enhancement compared to just including the neutral channel alone.

%%%%%%%%%%%%%%%%%%%%%%%%%%%%%%%%%%%%%%%%%%%%%%%%%%%%%%%%%%%%%%%%%%%%%%%%%%%
\subsection{Derivation of the reaction density $\boldsymbol{\gamma(T)}$} \label{ApE}
%%%%%%%%%%%%%%%%%%%%%%%%%%%%%%%%%%%%%%%%%%%%%%%%%%%%%%%%%%%%%%%%%%%%%%%%%%%
In this appendix, we derive the reaction density $\gamma(T)$. Specifically, we show explicitly how eq.~\eqref{abs} follows from eq.~\eqref{eq:gamma} \& eq.~\eqref{abc}. Since production is dominated near threshold in the Boltzmann suppressed freeze-in limit $T\ll m$, we can parameterize the center of mass energy as $s\simeq 4m_\Psi^2+\varepsilon$ with $\varepsilon\ll 4m_\Psi^2$. It follows that
\beq
    \beta\simeq\sqrt{\frac{\varepsilon}{4m_\Psi^2}}\ll1.
\eeq
The product $(s-4m_\Psi^2)  \sqrt{s}  \sigma(s)$ which appears in the square brackets of eq.~\eqref{eq:gamma} can be approximated near threshold as
\beq
    (s - 4  m_\Psi^2)  \sqrt{s}  \sigma(s) \simeq  \sum_{q,q'}\left(\frac{\mathcal  C_\gamma^{qq'} + \mathcal C_Z^{qq'} +\mathcal C_W^{qq'}}{96\pi m_\Psi^{2}} \right) \varepsilon^{3/2}.
\eeq
Since $s\sim m_\Psi\gg T$ we can use the large value approximation of the Bessel function $K_1(z) \simeq \sqrt{\frac{\pi}{2z}}  e^{-z}$ for $z\gg1$, and working to first order in $\varepsilon$ we obtain
\beq
    K_1\left(\frac{\sqrt{s}}{T}\right) \simeq \sqrt{\frac{T\pi}{4m_\Psi}} e^{-\frac{2m_\Psi}{T}} e^{-\frac{\varepsilon}{4m_\Psi T}}~.
\eeq
Thus, we can rewrite the reaction density as an integral over $\varepsilon$ as follows
\beq
    \gamma(T) \simeq  \sum_{q,q'}\frac{T}{32\pi^4} \left( \frac{\mathcal C_\gamma^{qq'} + \mathcal C_Z^{qq'}+ \mathcal C_W^{qq'}}{96\pi m_\Psi^{2}} \right) \sqrt{\frac{\pi T}{4m_\Psi}} e^{-\frac{2m_\Psi}{T}} \int_0^\infty {\rm d}\varepsilon ~ \varepsilon^{3/2} e^{-\frac{\varepsilon}{4m_\Psi T}}.
\eeq
The $\varepsilon$ integral evaluates to $24\sqrt{\pi} m_\Psi^{5/2}T^{5/2}$ and thus we obtain the reaction for $m_\Psi \gg T$
\beq
    \gamma(T) \simeq \sum_{q,q'}\left(\frac{ \mathcal C_\gamma^{qq'} + \mathcal C_Z^{qq'}+ \mathcal C_W^{qq'}}{256 \pi^4} \right)T^{4} e^{-2m_\Psi/T}.
\eeq
This can be specialized to the neutral or charged components by the appropriate choice of $\mathcal C$.

%%%%%%%%%%%%%%%%%%%%%%%%%%%%%%%%%%%%%%%%%%
\subsection{Indirect Detection}
\label{ApF}
\vspace*{-2mm}
%%%%%%%%%%%%%%%%%%%%%%%%%%%%%%%%%%%%%%%%%%%%
In regions of high density, such as the galactic center, the dark matter $\Psi^0$ can undergo pair annihilation into Standard Model particles at a late time.  While this can potentially lead to indirect detection signals. In the high mass limit $m_\Psi\gg m_t$ the leading annihilation route is into $W^+W^-$ via a $t$-channel diagram. The $\Psi^0$ are non-relativistic at late time, with velocities of the order of $\sim 10^{-3}$ in virialized environments such as galactic halos. As a result, the long range exchange of gauge bosons between $\Psi^0$ states induces an attractive potential and enhances the tree-level $\Psi^0$ pair-annihilation cross-sections due to the Sommerfeld effect; see, e.g.~\cite{Hisano:2004ds, Feng:2010zp}. 

For brevity, we restrict our attention to the minimal model. Defining the two-body state 
\beq
    \psi_i = \{\bar{\Psi}^0 \Psi^0, \Psi^+ \Psi^-\},
\eeq
the form of the interaction potential $V_{ij}$ corresponding to the Lagrangian of eq.~\eqref{eq:4} is given by
\beq
    V_{ij}(r) \simeq
    \begin{pmatrix}
        0 &\qquad -\frac{1}{r}\alpha_2 e^{-m_W r} \\
        -\frac{1}{r}\alpha_2 e^{-m_W r} &\qquad 2 \Delta - \frac{1}{r} \alpha_{\rm em} - \frac{\alpha_2 (1 - 2\text{sin}^2 \theta_W)^2}{4r\text{cos}^2 \theta_W} e^{-m_Z r}
    \end{pmatrix}~,
\eeq
where $\alpha_2 \equiv g^2/4\pi$. The Sommerfeld factors $s_i$ can be determined by solving the coupled non-relativistic $s$-wave Schrödinger equations for $\psi_i(r)$ with the boundary conditions $\psi_i(\infty) = \{e^{i k r}, 0\}$, using the variable phase method outlined in e.g.~\cite{Asadi:2016ybp}. This leads to the Sommerfeld enhanced $s$-wave cross-sections
\beq
    \langle \sigma v \rangle = \sum_{i, j} \sum_{G, G'} s_i \Gamma^{GG'}_{ij} s_j^*~,
    \label{52}
\eeq
where $\Gamma^{GG'}_{ij}$ denotes the tree-level annihilation rates to gauge bosons $G$, $G'$ and can be written as
\begin{align}
\begin{split}
    \Gamma^{WW}_{ij} &\simeq \frac{\pi \alpha_2^2}{16 M^2} \begin{pmatrix}
        2 & \sqrt{2} \\
        \sqrt{2} & 4
    \end{pmatrix}~, 
    \hspace{3cm}
    \Gamma_{ij}^{ZZ} \simeq \frac{\pi \alpha_2^2 \left(1 - 2 \text{sin}^2 \theta_W \right)^4}{16M^2 \text{cos}^4 \theta_W }\begin{pmatrix}
        1 & 0 \\
        0 & 0
    \end{pmatrix}~, \\[12pt]
    \Gamma_{ij}^{\gamma Z} &\simeq \frac{\pi \alpha_{\rm em} \alpha_2 (1 -2 \text{sin}^2 \theta_W)^2}{M^2} \begin{pmatrix}
        1 & 0 \\
        0 & 0
    \end{pmatrix}~, 
    \qquad\qquad
    \Gamma_{ij}^{\gamma \gamma} \simeq \frac{\pi \alpha_{\rm em}^2}{M^2} \begin{pmatrix}
        1 & 0 \\
        0 & 0
    \end{pmatrix}~.
\end{split}
\label{53}
\end{align}

%%%%%%%%%%%%%%%%%%%%%%%%%%%%%%%%%%%%%%%%%%%%%%
\begin{figure}[t!]
    \includegraphics[width=0.5\columnwidth]{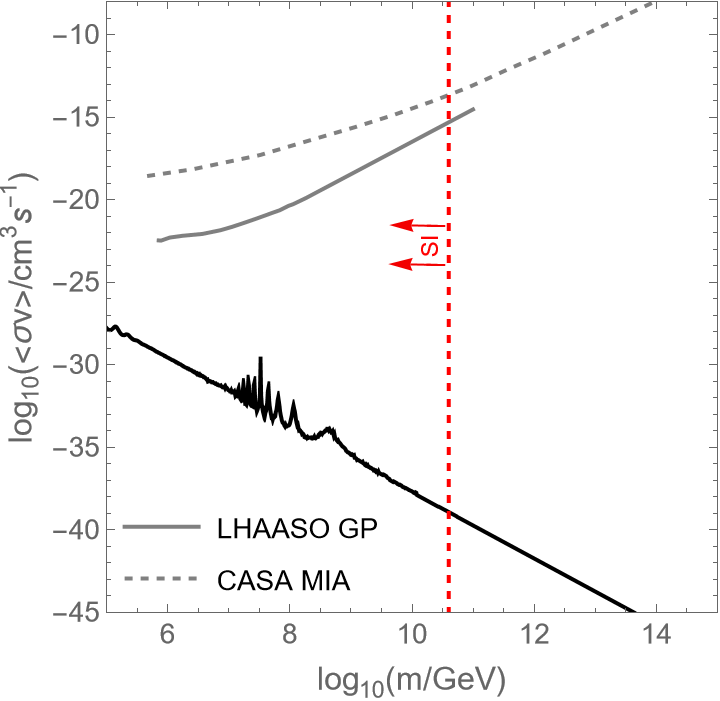}\vspace{-2pt}
    \caption{Indirect detection bounds on the vector-like model. The total annihilation cross-section (for $v=10^{-3}$) is shown as the black solid curve. We also present the leading constraints, namely, LHAASO galactic plane bounds \cite{Boehm:2025qro, LHAASO:2024lnz} and CASA-MIA~\cite{CASA-MIA:1997tns}, assuming 100\% annihilations to $W^+W^-$. The red dotted line marks the mass below which the (SI) scattering cross-section is excluded by direct detection.  Note the main sensitivity of indirect detection experiments lies within the mass range that is already excluded by direct detection limits.}
    \label{fig:3}
\end{figure}
%%%%%%%%%%%%%%%%%%%%%%%%%%%%%%%%%%%%%%%%%%%%%%

We use eqs.~\eqref{52} and \eqref{53} to calculate the Sommerfeld enhanced cross-section by summing over annihilation channels of dark matter into gauge boson pairs:
\beq
    GG' = \{W^+W^-,~ZZ,~\gamma Z,~\gamma \gamma\}.
\eeq
The total annihilation cross-section $\bar \Psi^0\Psi^0\rightarrow GG'$ is shown as the black solid curve in Fig.~\ref{fig:3}. We note that the dominant annihilation channel is $W^+W^-$ with  other channels contributing at subleading order, thus
\beq
    \langle\sigma v\rangle_{\rm tot}\approx \langle\sigma v\rangle_{WW} .
\eeq
One can compare the annihilation cross-section with experimental bounds. Figure \ref{fig:3} shows the leading indirect detection constraints: showing constraints \cite{Boehm:2025qro} (`Max' limits)  derived from LHAASO's galactic plane continuum $\gamma$-rays \cite{LHAASO:2024lnz} and $\gamma$-ray limits  based on CASA-MIA \cite{CASA-MIA:1997tns} (also cf.~\cite{Hiroshima:2025jyz})  We apply limits assuming 100\% annihilation to $W^+W^-$ pairs, since this matches the full annihilation spectrum to good approximation. Inspecting Fig.~\ref{fig:3} we conclude that even the strongest indirect detection limits due to LHAASO are not constraining over the parameter space of interest (i.e.~over the parameter space not already excluded by direct detection).

\end{document}